\documentclass[aps,prb,twocolumn,floats,showpacs,superscriptaddress]{revtex4-1}
\usepackage{graphicx,epsfig}
\usepackage{times,bbm}
\usepackage{graphics,dcolumn,bm,float}
\usepackage{amssymb,amsmath,rotate,color}
\usepackage[title,titletoc,toc]{appendix}
\usepackage{mathtools}
\usepackage{booktabs}
\usepackage{tcolorbox}

\usepackage[pagebackref=false,colorlinks,linkcolor=magenta,citecolor=cyan,urlcolor=magenta]{hyperref}

\begin{document}
\unitlength 1 cm
\newcommand{\be}{\begin{equation}}
\newcommand{\ee}{\end{equation}}
\newcommand{\nn}{\nonumber}
\newcommand{\vk}{\vec k}
\newcommand{\vp}{\vec p}
\newcommand{\vq}{\vec q}
\newcommand{\vkp}{\vec {k'}}
\newcommand{\vpp}{\vec {p'}}
\newcommand{\vqp}{\vec {q'}}
\newcommand{\bk}{{\vec k}}
\newcommand{\bp}{{\bf p}}
\newcommand{\bq}{{\bf q}}
\newcommand{\br}{{\bf r}}
\newcommand{\bR}{{\bf R}}
\newcommand{\up}{\uparrow}
\newcommand{\down}{\downarrow}
\newcommand{\cdag}{c^{\dagger}}
\newcommand{\hlt}[1]{\textcolor{red}{#1}}
\newcommand{\ba}{\begin{align}}
\newcommand{\ea}{\end{align}}
\newcommand{\la}{\langle}
\newcommand{\ra}{\rangle}

 \title{Topological phase transiton of anisotropic XY model with Dzyaloshinskii-Moriya interaction}
   \author{T. Farajollahpour}
   \affiliation{Department of Physics, Sharif University of Technology, Tehran 11155-9161, Iran}

 \author{S. A. Jafari}
 \email{akbar.jafari@gmail.com}
 \affiliation{Department of Physics, Sharif University of Technology, Tehran 11155-9161, Iran}
 \affiliation{Center of excellence for Complex Systems and Condensed Matter (CSCM), Sharif University of Technology, Tehran 1458889694, Iran}

\begin{abstract}
Within the real space renormalization group we obtain the phase portrait of the
anisotropic quantum XY model on square lattice in presence of Dzyaloshinskii-Moriya (DM) interaction.
The model is characterized by two parameters, $\lambda$ corresponding to XY anisotropy, and $D$ 
corresponding to the strength of DM interaction. The flow portrait of the model is governed by 
two global Ising-Kitaev attractors at $(\lambda=\pm1,D=0)$ and a repeller line, $\lambda=0$.
Renormalization flow of concurrence suggests that the $\lambda=0$ line corresponds to a
topological phase transition. 
The gap starts at zero on this repeller line corresponding to super-fluid
phase of underlying bosons; and flows towards a finite value at the Ising-Kitaev points. At these two fixed points 
the spin fields become purely classical, and hence the resulting Ising degeneracy can be interpreted as topological 
degeneracy of dual degrees of freedom. 
The state of affairs at the Ising-Kitaev fixed point is consistent with the picture of a p-wave pairing of 
strength $\lambda$ of Jordan-Wigner fermions coupled with Chern-Simons gauge fields. 
\end{abstract}

\maketitle

\section{introduction}
The two-dimensional classical (vector) XY model is a paradigm for the celebrated Berezinskii-Kosterlitz-Thouless (BKT) 
transition upon which the phase coherence of an underlying super-fluid is lost by the proliferation of topological
excitations known as vortices~\cite{Berezinskii1971,KT1973,FradkinBook}.
Quantum version of this model was initially proposed by Matsubara and Matsuda as a lattice model 
to understand the liquid helium~\cite{Matsubara1956}. 
Since then there has been tremendous studies of the the two-dimensional quantum XY (2DQXY) model. 
Berezinskii used the term "anisotropic planar magnetic substances" to refer to the quantum XY model~\cite{Berezinskii1972}.
The isotropic limit of the XY model refers to the situation where $\sigma^x\sigma^x$ and $\sigma^y\sigma^y$
couplings have equal strength. This is the isotropic limit of the XY model. 
Oitma and Betts found that the ground state of this model has finite transverse magnetization~\cite{Betts1978}.
The exact diagonalization study of Tang on the anti-ferromagnetic XY model found isotropic staggered
magnetization in the XY plane~\cite{Tang1988}. Drzewinsky and Sznajd used a block-spin renormalization
group at finite temperatures to find a BKT transition temperature in this system~\cite{Sznajd1989}. 
The BKT transition for the 2DQXY was confirmed in quantum Monte Carlo studies~\cite{Ding1990,Ding1992,Ying1998,Harada}.
The critical exponents extracted from the quantum Monte Carlo study of Ding and coworkers suggested that
2DQXY belongs to the same universality class as the classical (vector) XY model~\cite{Ding1992}.

An equivalent way of thinking about 2DQXY model is in terms of hardcore bosons~\cite{Rigol2012}.
This bosonic language is particularly convenient for the study of super-fluid transition 
measured by super-fluid density, $\rho_s$, which in the spin language
corresponds to spin-stiffness~\cite{SachdevBook}.
In the bosonic language for a system with filling fraction $n$ at the classical level
the zero temperature super-fluid-density is given by $\rho_s^{\rm cl}\propto n(1-n)$. 
Quantum fluctuations enhance the above stiffness by few percent~\cite{Sandvik1999}. 
The emerging picture is that the zero temperature phase of the isotropic 2DQXY is that of a super-fluid.
Indeed in a remarkable paper a much stronger version of this for all spins and for all dimensions higher
than one was proven by Kennedy, Lieb and Shastry~\cite{Shastry}.

Extensions of the isotropic 2DQXY model are also very interesting. 
Dekeyser and coworkers employed the quantum renormalization group method to suggest that 
extending the 2DQXY by an $\sigma^z\sigma^z$ Ising term gives a very simple picture that the greater of 
Ising and XY exchange interaction dominates the low-energy phase~\cite{Dekeyser1977}.
Such an Ising exchange is equivalent to interaction among bosons. Placing this model on
triangular lattice~\cite{Heidarian} sets a very interesting competition between the Mott localization,
geometric frustration and super-fluidity of hardcore bosons where a diagonal
solid order emerges at strong interactions~\cite{Arun} and remains stable for arbitrary large values of
interaction~\cite{Heidarian}. This can be a possible explanation for the super-solid phase of helium~\cite{Supersolid}.
Another possible extension is by plaquette interactions in presence of an external field where the
four-site terms encourage valence bond solid~\cite{Melko2004}.
Allowing for bond-disorder in the 2DQXY model enhances the amplitude of 
zero-point phase fluctuations giving rise to vanishing of the spin-stiffness which then
turns the ground state into spin liquid~\cite{Gawiec1996}.

In addition to the above bosonic picture of the 2DQXY model and its extensions, 
there is also fermionic picture which is based on a Jordan-Wigner transformation.
In this approach the spin system is mapped into a Chern-Simons (CS)
theory coupled with spin-$1/2$ fermions~\cite{Fradkin1989,Fradkin1994,FradkinBook}.
This mapping is quite general and applies to larger family of spin systems than the 2DQXY
on any bipartite lattice~\cite{Wang1991}. This approach is quite powerful, and is used to
relate the $1/3$ magnetization plateau of the regime of XY anisotropy to a bosonic fractional 
Laughlin state with filling fraction $1/2$~\cite{Fradkin2014}.

In this paper we extend the {\em anisotropic} 2DQXY model 
by adding a Dzyaloshinskii-Moriya (DM) interaction of strength $D$. 
We consider a planar anisotropy $\lambda$ that makes the exchange in $x$ and $y$ directions different. 
On top of that we add a DM interaction between the planar components 
of the spin. We employ the block-spin renormalization group (BSRG)
to study the phase transitions of this model. We construct a phase portrait of the model
from our BSRG equations. We find that there are two global attractors that attract the 
flow to gapped states which correspond to Ising phases polarized along $x$ or $y$ directions~\cite{Tang1988}. These two are separated by a gap-closing and hence 
should correspond to topologically non-trivial phases, similar to its one-dimensional counterpart~\cite{SAJ2017}.
We corroborate the topological nature of this quantum phase transition with the calculation of concurrence. 
The whole $\lambda=0$ line in the plane of $\lambda$ and $D$ will be a gapless repeller which is unstable 
with respect to smallest anisotropy $\lambda$ (irrespective of the sign of $\lambda$). This is reminiscent of the
pairing instability in a gapless system of Jordan-Winger fermions~\cite{SAJ2017}
which from the exact solution of the one-dimensional problem can be interpreted as the
p-wave pairing interaction. Indeed such a p-wave pairing resulting from the anisotropy
$\lambda$ can be obtained from the study of equivalent fermions coupled to Chern-Simons
gauge fields on the honeycomb lattice~\cite{Kamenov2017}. 

The paper is organized as follows. In Sec.~\ref{XYRG} the XY model in the presence of DM 
interaction has been considered. The effective Hamiltonian of the system for the renormalized 
coupling constant and anisotropic parameters is obtained. In the Sec.~\ref{phases} we present the details 
of the phase diagram. The discussions and results are presented in Sec.~\ref{Discuss}.  

\section{Model and method} \label{XYRG}
\begin{figure}[t]
	\centering
	\includegraphics[width =0.800 \linewidth]{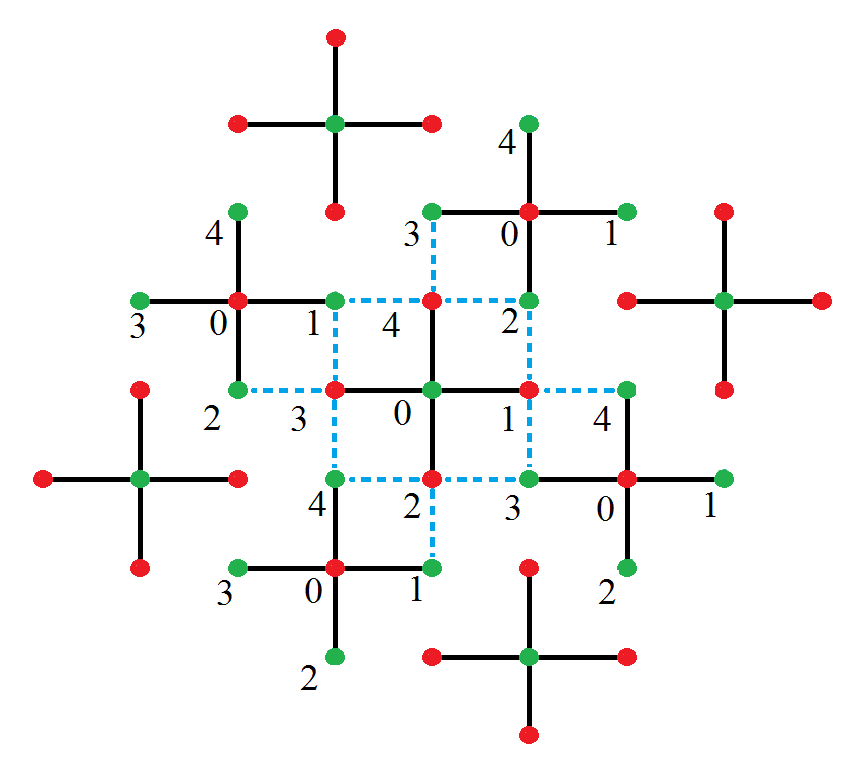}
	\caption{(Color online) The selected cluster in a square lattice where the 
		dashed lines shows the block-block interactions.}
	\label{Cluster}
\end{figure}

The Hamiltonian of XY model on a 2D square lattice in the presence of DM interaction with 
N$\times$N sites can be written as,
\begin{align}
H(J,\lambda,D)& = \nonumber\\
& J\sum_{i=1}^{N} \sum_{j=1}^{N} [(1+\lambda) \left(\sigma_{i,j}^x\sigma_{i+1,j}^x + \sigma_{i,j}^x\sigma_{i,j+1}^x\right) \nonumber\\
&+(1-\lambda)(\sigma_{i,j}^y\sigma_{i+1,j}^y + \sigma_{i,j}^y\sigma_{i,j+1}^y) \nonumber\\
& +D (\sigma_{i,j}^x\sigma_{i+1,j}^y - \sigma_{i,j}^y\sigma_{i+1,j}^x ) \nonumber\\
&+ D (\sigma_{i,j}^x\sigma_{i,j+1}^y - \sigma_{i,j}^y\sigma_{i,j+1}^x )]
\label{HXY}
\end{align}
where $J>0$ is the exchange coupling, $\lambda$ is anisotropy parameter, 
$D$ is the DM interaction term and $\sigma_i^n ~(n=x,y,z)$ are Pauli matrices 
at site $i$. 
The basic idea of block-spin renormalization method is to partition the lattice into
clusters. Then if the cluster allows for a Kramers doublet ground states, the
fluctuations between such doublet can be captured with an effective (coarse grained)
spin variable~\cite{Delgado, Delgado1}. 

\subsection{Block spin RG equations}
\begin{figure}[t]
	\centering
	\includegraphics[width = 1.00 \linewidth]{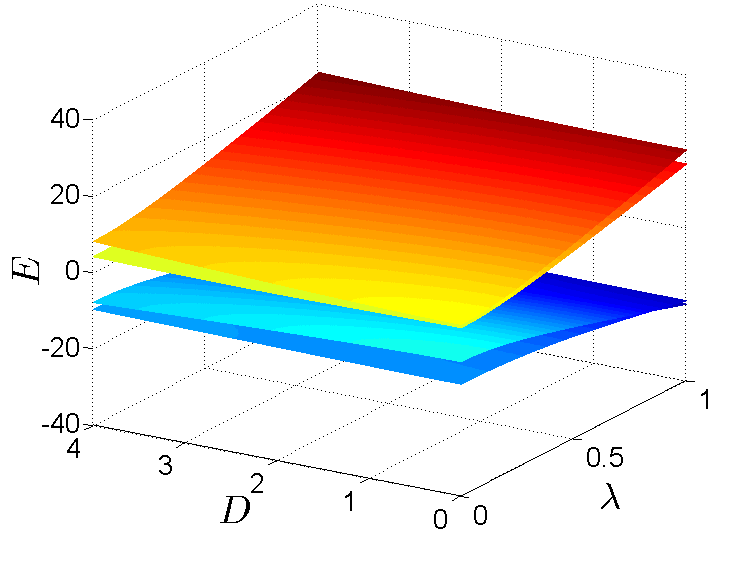}
	\caption{(Color online) (a) The bands plots of selected five-site cluster in terms of 
		$\lambda$ and $J$ when $D = 0$. (b) The plots of bands in nonzero DM interaction 
		at $J=1$. }
	\label{bandsXY}
\end{figure}

To study the ground state phases of the
above Hamiltonian, we partition the square lattice into blocks of five sites 
as depicted in Fig.~\ref{Cluster}. Out of the five sites in the cluster, four are
from one sub-lattice and one is from the other sub-lattice. For interactions involving
the Ising term of the form $\sigma^z\sigma^z$ such a sub-lattice imbalance erroneously 
biases the ground state towards the wrong ground state total spin. This is due to Lieb-Mattis theorem
for the Hubbard and Heisenberg family of models. However for XY family where 
the only conserved charge is $\zeta=\prod_j\sigma^z_j$~\cite{SAJ2017} where $j$ runts over the 
whole lattice, this sub-lattice asymmetry does not destroy the doublet structure 
of the ground state and we still get a doublet of ground states each belonging to 
$\zeta=\pm 1$ sectors. The conserved charge $\zeta$ already breaks the $2^5$ dimensional
Hilbert space into two sectors, each of dimension $16$. States in each sector
are in one-to-one correspondence in the above two sectors. These two sectors are
mapped to each other by replacing the role of $\up$ and $\down$ spins. 
The clusters in Fig.~\ref{Cluster} have further four-fold rotational symmetry.
This allows to use standard methods of group theory~\cite{group} to further reduce 
the $16$ dimensional space corresponding to a given $\zeta$. The details of the straightforward but
lengthily algebra is given in the appendix. The sector that contains the ground state
is a $6\times 6$ dimensional space which can be diagonalized to give the set of
eigenvalues depicted in Fig.~\ref{bandsXY} in the parameter space of $D,\lambda$. 

The ground state energy in both $\zeta=\pm 1$ sectors is
\begin{eqnarray}
e_0 =-2J \sqrt{5 (1 +D^2)+ 5 \lambda^2 + \eta} ,
\label{Eig}
\end{eqnarray}
where 
\begin{eqnarray}
\eta = \sqrt{\lambda^4 + 34 \lambda^2 (1 + D^2) + (1 + D^2)^2}
\end{eqnarray}
and the ground state eigen-vector in the $\zeta=+1$ sector is,
\begin{align}
|\phi_+\rangle=&\gamma_1 |\downarrow \downarrow \downarrow \downarrow \downarrow \rangle
+ \gamma_2 |\uparrow \uparrow \uparrow \uparrow \downarrow\rangle+ 
\gamma_3 (| \uparrow \uparrow \uparrow\downarrow \uparrow \rangle 
+ | \uparrow \uparrow\downarrow \uparrow \uparrow \rangle \nonumber\\  
&+| \uparrow\downarrow \uparrow \uparrow \uparrow \rangle +
| \downarrow\uparrow \uparrow \uparrow \uparrow \rangle)+ 
\gamma_4 (|\uparrow \downarrow\downarrow\downarrow \uparrow \rangle + 
|\downarrow \downarrow\downarrow\uparrow \uparrow \rangle \nonumber\\
&+ |\downarrow \uparrow\downarrow\downarrow \uparrow \rangle 
+ |\downarrow \downarrow\uparrow\downarrow \uparrow \rangle
)+ \frac{\sqrt{2}}{2} (|\uparrow \downarrow\downarrow\uparrow \downarrow \rangle
+ |\uparrow \uparrow\downarrow\downarrow \downarrow \rangle
\nonumber\\
&+ |\downarrow \downarrow\uparrow\uparrow \downarrow \rangle
+ |\downarrow \uparrow\uparrow\downarrow \downarrow \rangle
+|\uparrow \downarrow\uparrow\downarrow \downarrow \rangle +
|\downarrow \uparrow\downarrow\uparrow \downarrow \rangle).
\label{Eigv1}
\end{align}
The ground state in $\zeta=-1$ sector is simply obtained by the spin-flip
transformation of the above ground state, $\up \leftrightarrow \down$. 
\begin{align}
|\phi_- \rangle=&\gamma_1 |\uparrow \uparrow \uparrow \uparrow \uparrow \rangle
+ \gamma_2 |\downarrow \downarrow \downarrow \downarrow \uparrow\rangle+ 
\gamma_3 (| \downarrow \downarrow \downarrow\uparrow \downarrow \rangle 
+ | \downarrow \downarrow\uparrow \downarrow \downarrow \rangle \nonumber\\  
&+| \downarrow\uparrow \downarrow \downarrow \downarrow \rangle +
| \uparrow\downarrow \downarrow \downarrow \downarrow \rangle)+ 
\gamma_4 (|\downarrow \uparrow\uparrow\uparrow \downarrow \rangle + 
|\uparrow \uparrow\uparrow\downarrow \downarrow \rangle \nonumber\\
&+ |\uparrow \downarrow\uparrow\uparrow \downarrow \rangle 
+ |\uparrow \uparrow\downarrow\uparrow \downarrow \rangle
)+ \frac{\sqrt{2}}{2} (|\downarrow \uparrow\uparrow\downarrow \uparrow \rangle
+ |\downarrow \downarrow\uparrow\uparrow \uparrow \rangle
\nonumber\\
&+ |\uparrow \uparrow\downarrow\downarrow \uparrow \rangle
+ |\uparrow \downarrow\downarrow\uparrow \uparrow \rangle
+|\downarrow \uparrow\downarrow\uparrow \uparrow \rangle +
|\uparrow \downarrow\uparrow\downarrow \uparrow \rangle)
\label{eigv2}
\end{align}
where the coefficients are obtained as,
\begin{align}
& \gamma_1 = \frac{6 \sqrt{2} \lambda(1 + iD)}{5 (1+D^2) + \lambda^2 + \eta} \nonumber\\
& \gamma_2 = \frac{\sqrt{2}(1+iD)(-1 + \lambda^2 - D^2+ \eta)}{\lambda (5 (1+D^2) + \lambda^2  +\eta)} \nonumber\\
& \gamma_3 = \frac{(-1 - 5\lambda^2-D^2 + \eta)\sqrt{5 (1+D^2) + 5\lambda^2 + \eta} }{4\sqrt{2} \lambda (-1 -D^2+ \lambda^2)} \nonumber\\
& \gamma_4 = -\frac{3i(-i+D)\sqrt{5 (1+D^2) + 5\lambda^2 + \eta}}{{5 (1+D^2)+  \lambda^2 + \eta}}.
\end{align}

As illustrated in Fig.~\ref{bandsXY} the presence of DM interaction 
will not generate any band crossing and the ground state remains stable 
with respect to change of anisotropy parameters and DM interactions. 
The relation between Hamiltonian~\eqref{HXY} and the coarse-grained
effective Hamiltonian is formally given as,
\begin{eqnarray}
H^{\rm eff} = T^\dagger_0 H T_0,
\label{Heff}
\end{eqnarray}
where the projection operator $T_0$ basically assigns a new coarse grained spins $\Uparrow,\Downarrow$
to the two degenerate (Kramers double) ground states $|\phi_+\rangle$ and $|\phi_-\rangle$ in the
$\zeta=\pm 1$ sectors:
\begin{eqnarray}
T_0 = | \phi_+ \rangle \langle \Uparrow| + |\phi_- \rangle \langle  \Downarrow|.
\label{T0}
\end{eqnarray}
The Pauli matrix $\sigma^z_j$ can not change the charge $\zeta=\pm 1$ and therefore
in the space composed of doublet of $|\phi_\pm\rangle$, the action of each $\sigma^z_j$
contributes to the formation of a coarse grained $\sigma'^z$ for the whole cluster.
Similarly each $\sigma^{x(y)}_j$ flips one of the spins, thereby flipping the sign
of $\zeta$ and can be interpreted as flipping the coarse grained spins $|\Uparrow\rangle$ and
$|\Downarrow\rangle$ which can then be represented by $\sigma'^{x(y)}$ in the space of 
coarse grained spins. In this process some coefficients from the ground state wave function
will be collected. For the coupling of neighboring coarse-grained spins one needs to 
collect interactions on the bonds connecting a cluster to its neighbors~\cite{SAJ2017}. 
The effective Hamiltonian in the $n$'th step of the above process will be
\begin{align}
H(J_n,\lambda_n,D_n) & = \nonumber\\
& J_n\sum_{r=1}^{N/5} \sum_{s=1}^{N/5} [(1+\lambda_n) (\sigma_{r,s}^x\sigma_{r+1,s}^x + \sigma_{r,s}^x\sigma_{r,s+1}^x) \nonumber\\
& +(1-\lambda_n)(\sigma_{r,s}^y\sigma_{r+1,s}^y + \sigma_{r,s}^y\sigma_{r,s+1}^y) \nonumber\\
& + D_n (\sigma_{r,s}^x\sigma_{r+1,s}^y - \sigma_{r,s}^y\sigma_{r+1,s}^x ) \nonumber\\
&+ D_n (\sigma_{r,s}^x\sigma_{r,s+1}^y - \sigma_{r,s}^y\sigma_{r,s+1}^x )],
\end{align}
where the BSRG transformation connecting two consecutive steps becomes,
\begin{eqnarray}
 J_{n+1} = \frac{\alpha^2 + \xi^2 + \left(\alpha^2 - \xi^2 \right)\lambda_n }{2}J_n \nonumber\\
 \lambda_{n+1} = \frac{\alpha^2 - \xi^2 + \left(\alpha^2 + \xi^2 \right)\lambda_n}{\alpha^2 + \xi^2 + \left(\alpha^2 - \xi^2 \right)\lambda_n} \nonumber\\
D_{n+1} =\frac{2\alpha \xi}{\alpha^2 + \xi^2 + \left( \alpha^2 - \xi^2 \right)\lambda_n}D_n.
\end{eqnarray}
The coefficients are given by,
\begin{align}
 \alpha=\frac{1}{\mathcal{N}^2}&[ \gamma_1\gamma_3^* + \gamma_3\gamma_1^* + \gamma_2\gamma_4^* + \gamma_4\gamma_2^* \nonumber\\
 &\frac{3\sqrt{2}}{2}(\gamma_3 + \gamma_3^*+ \gamma_4 + \gamma_4^*) ]
  \nonumber\\
 \xi = \frac{1}{\mathcal{N}^2}&[ \gamma_1\gamma_3^* + \gamma_3\gamma_1^* - \gamma_2\gamma_4^* - \gamma_4\gamma_2^* \nonumber\\
 &\frac{3\sqrt{2}}{2}(-\gamma_3 - \gamma_3^*+ \gamma_4 + \gamma_4^*) ]\nn
\end{align}
with
\begin{eqnarray}
\mathcal{N} = \sqrt{|\gamma_1|^2 + |\gamma_2|^2 +4|\gamma_3|^2 +4|\gamma_4|^2 +3}.\nn
\end{eqnarray}
The $\gamma_i^*$ is the conjugate of $\gamma_i$ and $|\gamma_i|^2 = \gamma_i \gamma_i^*$. 

\subsection{Entanglement analysis}\label{XYentan}
Among various tools, entanglement is standard tool for the diagnosis of the
phase transitions in general~\cite{reventanglement,Kar2007,Kar2008,Langari2004,song2013monogamy,song2014,Ma2011,liu2016renormalization} and
change of {\em topology}~\cite{kitaev,Balents,SCI2017} in particular. 
Therefore let us use this tool in our five-site problem and to study its 
evolution under the RG transformations obtained above. Let us consider
the entanglement between two spins in the corners of each blocks. 
For this purpose we first calculate the reduced density matrix between every 
two pairs of neighboring sites, namely, $\rho_{12}$, $\rho_{13}$, $\rho_{14}$, 
$\rho_{23}$, $\rho_{24}$ and $\rho_{34}$ which involves tracing out all the rest 
of degrees of freedom~\cite{bennett,concurrence}. To construct the $4\times 4$
matrix $\rho_{ij}$ ($i,j$ are site indices, not matrix indices) one first constructs the full density matrix,
\begin{eqnarray}
\rho = |\psi_0 \rangle \langle \psi_0|
\end{eqnarray}
where the $|\psi_0 \rangle$ can be any of the Kramers doublet degenerate ground states $|\phi_\pm\rangle$
and then traces all sites except for sites $i,j$. 
We then form a matrix $\rho_{ij} \tilde{\rho}_{ij}$ where 
$\tilde{\rho}_{ij} = \left(\sigma^y_i \otimes  \sigma^y_j\right) \rho^*_{ij} \left(\sigma^y_i \otimes  \sigma^y_j\right)$
and number its four eigenvalues such that $\lambda_{ij,m},~(m=1,2,3,4)$
such that $\lambda_{ij,4} > \lambda_{ij,3}> \lambda_{ij,2} >\lambda_{ij,1}$. 
From these eigenvalues we then evaluate the bipartite concurrence defined by~\cite{concurrence},
\begin{align}
C_{ij} =max \left(\sqrt{\lambda_{ij,4}}-\sqrt{\lambda_{ij,3}}-\sqrt{\lambda_{ij,2}}-\sqrt{\lambda_{ij,1}} , ~0 \right)\nn.
\end{align}
Then we construct a geometric mean of the concurrence between the above 6 pairs of sites as~\cite{concurrence},
\begin{align}
C_g=\sqrt[6]{C_{12} \times C_{13} \times C_{14} \times C_{23} \times C_{24} \times C_{34}}.
\end{align}
In the present case, by rotational symmetry all of the six density matrices are equal and given by 
\begin{figure}[t]
	\centering
	\includegraphics[width=0.95 \linewidth]{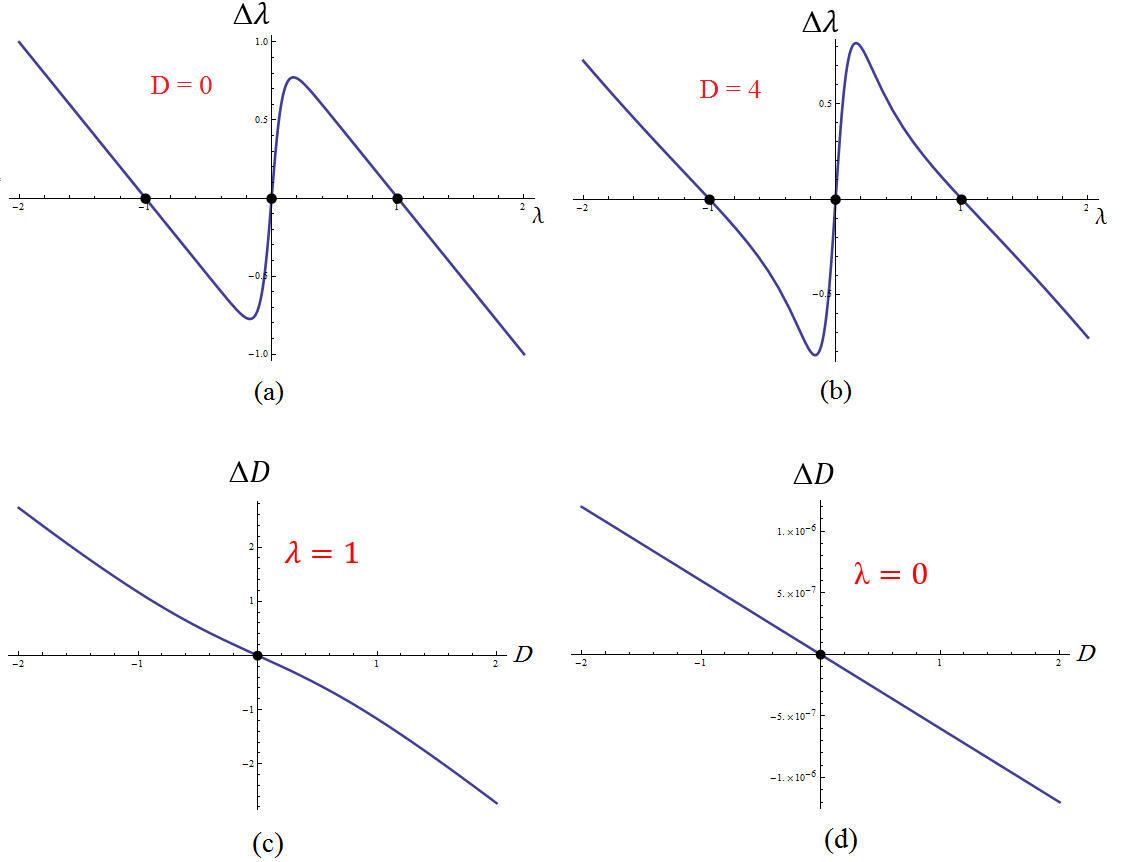}
	\caption{(Color online) Fixed points of the 2DQXY model with anisotropy ($\lambda$) 
	and DM interaction ($D$). In top (bottom) row we have fixed $D$ ($\lambda$) to study flow of $\lambda$ ($D$).
	There are two attractors at $\lambda=\pm 1$ and a repulsive fixed point at $\lambda=0$.
	The DM interaction has only one attractor at $D=0$. }
	\label{Fixed}
\end{figure}
\begin{widetext}
\begin{align}
\rho_{12} = \rho_{23} = \rho_{34} =&\rho_{13} =\rho_{14} = \rho_{24} = \nonumber\\
&=\frac{1}{\mathcal{N}^2}\left(
\begin{matrix}
\gamma_1\gamma_1^* + 2\gamma_4\gamma_4^* + \frac{1}{2} & 0 & 0 & \frac{\sqrt{2}}{2}(\gamma_1 + \gamma_2^*) + 2\gamma_4\gamma_3^*\\
0 &  \gamma_3\gamma_3^* + \gamma_4\gamma_4^* + 1 & \gamma_3\gamma_3^* + \gamma_4\gamma_4^* + 1 & 0 \\
0 &  \gamma_3\gamma_3^* + \gamma_4\gamma_4^* + 1 & \gamma_3\gamma_3^* + \gamma_4\gamma_4^* + 1 & 0 \\
 \frac{\sqrt{2}}{2}(\gamma_1^* + \gamma_2) + 2\gamma_4^*\gamma_3& 0 & 0 & \gamma_2\gamma_2^* + 2\gamma_3\gamma_3^* + \frac{1}{2} \\
\end{matrix}
\right)
\end{align}
The above matrix gives,
\begin{align}
C_g =\frac{\sqrt{\Gamma_2^2 + 2\Gamma_1\Gamma_3 + \Gamma_2^{*2} + 2\sqrt{\Lambda} \Re \Gamma_2}}{\sqrt{2}} - \frac{\sqrt{\Gamma_2^2 + 2\Gamma_1\Gamma_3 + \Gamma_2^{*2} - 2\sqrt{\Lambda}\Re \Gamma_2}}{\sqrt{2}}  -2\sqrt{\Gamma_4}
\end{align}
where
$\Lambda = \sqrt{\Gamma_2^2 + 4\Gamma_1\Gamma_3-2\Gamma_2\Gamma_2^* + \Gamma_2^{*2}} \nonumber$,
$\Gamma_1 =\frac{1}{\mathcal{N}^2} \left( \gamma_1\gamma_1^* + 2\gamma_4\gamma_4^* + \frac{1}{2} \right) \nonumber$, 
$\Gamma_2 =\frac{1}{\mathcal{N}^2} \left(\frac{\sqrt{2}}{2}(\gamma_1 + \gamma_2^*) + 2\gamma_4\gamma_3^* \right) \nonumber$, 
$\Gamma_3 = \frac{1}{\mathcal{N}^2} \left(  \gamma_3\gamma_3^* + \gamma_4\gamma_4^* + 1 \right) \nonumber$ and 
$\Gamma_4 = \frac{1}{\mathcal{N}^2} \left( \gamma_2\gamma_2^* + 2\gamma_3\gamma_3^* + \frac{1}{2} \right)$.
\end{widetext}
We will use the above formula in our analysis of the phase transitions of the 2DQXY with planar
anisotropy $\lambda$ and DM interaction $D$.

\section{Phase diagram of the model}\label{phases}
\subsection{Analysis of the phase portrait}
The standard method for analysis of the phase diagram of a model that depends on set of
parameters ${\mathbf R}$ is to study $\Delta{\mathbf R}_n\equiv{\mathbf R}_{n+1}-{\mathbf R}_n$ and
its dependence to the initial values ${\mathbf R}_0\equiv{\mathbf R}$~\cite{strogatz2014,SAJ2017}. In our problem
the parameters are given by ${\mathbf R}=(\lambda,D)$.  In Fig.~\ref{Fixed} we have presented two such cuts.
In the first row, for two fixed values of $D=0$ (left) and $D=4$ (right) we plot how $\Delta\lambda$ depends
on the initial value $\lambda$. As can be seen there are two fixed points. Repulsive fixed point at 
$\lambda_*^0=0$ and two attractors at $\lambda_*^\pm=\pm1$~\cite{SAJ2017}. These values do not change by replacing
$D=0$ with $D=4$. In the second row of Fig.~\ref{Fixed}, for two fixed values of $\lambda=0$ and $\lambda=1$
we have plotted how $\Delta D$ depends on the initial value $D$. As can be seen, independent of value of
$\lambda$, there is always an attractor at $D_*=0$: Slightly moving to the right (left) of $D_*=0$, gives a 
negative (positive) $\Delta D$ that returns $D$ to the attractor $D_*=0$.
Therefore the coupling $D$ is irrelevant and any Hamiltonian of the form~\eqref{HXY} with a non-zero $D$ 
in the long wave-length limit behaves similar to the $D_*=0$ and the DM interaction is renormalized
away in the infrared limit.

\begin{figure}[t]
\centering
\includegraphics[width =0.800 \linewidth]{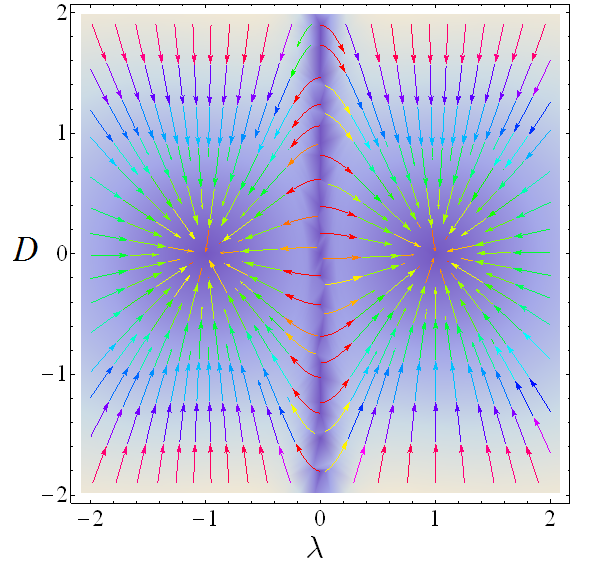}
\caption{(Color online) Phase portrait of the 2DQXY model with anisotropy $\lambda$ and
     DM interaction $D$. Two global attractors at $(\lambda_*^\pm,D_*)=(\pm1,0)$ along with repeller at infinity and
     a repulsive line $(\lambda_*^0,D)$ for every $D$, completely characterize the above RG flow profile. 
	}
\label{platform}
\end{figure}

Let us put the above picture in a global perspective in a plane composed of $\lambda$ and $D$. 
In Fig.~\ref{platform} we have provided a stream plot of the vectors $\Delta{\mathbf R}=(\Delta\lambda,\Delta D)$
as a function of the initial value ${\mathbf R}=(\lambda,D)$. 
As can be seen the fact that in Fig.~\ref{Fixed} the fixed point at $\lambda_*^\pm$ does not
depend on $D$ is reflected in Fig.~\ref{platform} as the fact that the two attractors at $(\lambda_*^\pm,D_*)=(\pm1,0)$
are globally attractive fixed points. However the fact that the repulsive fixed point $\lambda_*^0$
in Fig.~\ref{Fixed} does not depend on $\lambda$ is reflected in Fig.~\ref{platform} as a repulsive line.
The symmetry of the above phase portrait under $\lambda\to -\lambda$ is the direct manifestation
of the fact that Hamiltonian is invariant under $\sigma_j^x\to\sigma_j^y$, $\sigma_j^y\to -\sigma_j^x$ 
($\pi/2$ rotation around $z$ axis), $D\to D$ and $\lambda\to -\lambda$.

\subsection{Analysis of the gap}
\label{XYgap}
\begin{figure}[t]
	\centering
	\includegraphics[width=0.75 \linewidth]{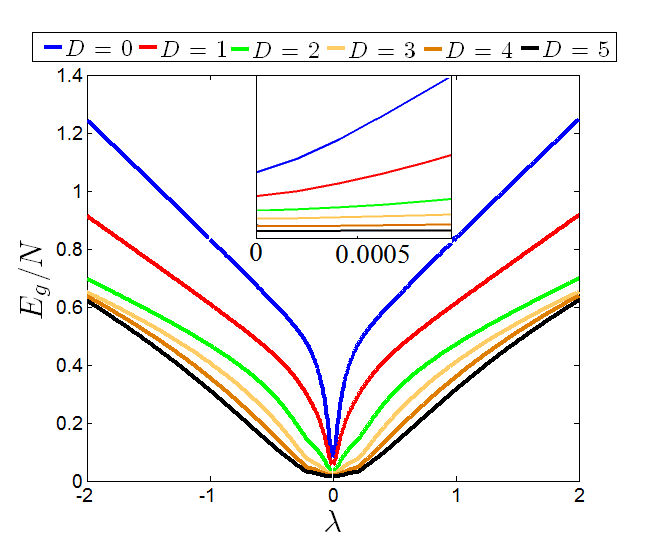}
	\caption{(Color online) Dependence of the gap on anisotropy $\lambda$ for various values of
	the DM interaction $D$ indicated in the legend.}
	\label{gapXY}
\end{figure}

So far our phase portraits in Figs.~\ref{Fixed} and~\ref{platform} indicate the irrelevance
of $D$ and a possible phase transition at $\lambda=0$ line. Let us see how does this manifest
itself in the spectral gap. 
The gap between the ground state and the first excited state is given by,
\begin{align}
 E_g^C &=\\
& 2J\left[\sqrt{5(1+D^2) + 5\lambda^2 + \eta }-\sqrt{5(1+D^2) + 5\lambda^2 - \eta }\right] \nn
\end{align}
The effect of RG flow on this quantity when it is iterated up to large enough RG steps to ensure 
machine precision convergence is plotted in Fig.~\ref{gapXY} for various values of the DM interaction $D$ indicated in the legend.
In this figure we plot the gap at the 8-th RG step (converged
within $10^{-5}$).
As can be seen for every value of $D$, the point $\lambda_*^0=0$ is the only gapless point, and
any non-zero value of $\lambda$, either positive or negative gives rise to a non-zero gap. The gap is
normalized per lattice site, and the natural unit of the gap is $J$.
The fact that for every value of $D$ we have a non-zero gap for $\lambda\ne 0$ agrees with
the existence of a line of fixed points $\lambda=\lambda_*^0$ in the $(\lambda,D)$ plane of Fig.~\ref{platform}.

\begin{figure}[b]
 	\centering
 	\includegraphics[width=0.75 \linewidth]{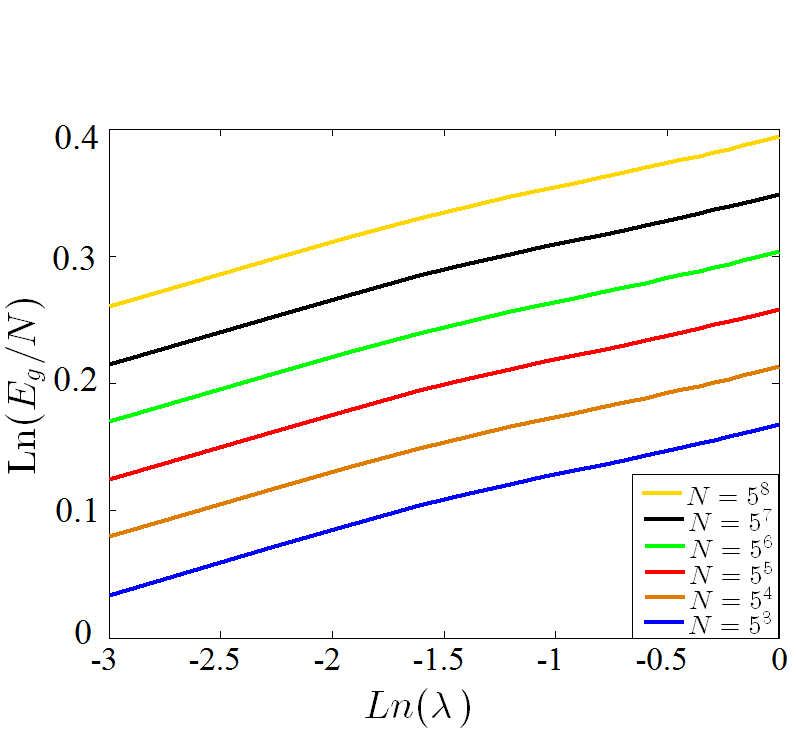}
 	\caption{(Color online) The power law behavior of gap 
 		in terms of anisotropy parameter $\lambda \neq 0$ for $D=0$.
		Different colors correspond to various RG steps as indicated in the legend.}
 	\label{LnEgn1}
\end{figure}

As can be seen in Fig.~\ref{gapXY} although for all values of $D$ the gap is a function of $\lambda$ that
vanishes at $\lambda=0$, but the way it vanishes depends on $D$ and is not universal. 
To extract these information, in Fig.~\ref{LnEgn1} we produce a log-log
plot of the gap versus $\lambda$ for $D=0$. Note that very small values of $\lambda\sim 10^{-3}$ are needed to 
extract the dependence of gap on $\lambda$. The linear dependence of the log-log plot suggests 
a perfect power-law dependence of the gap, $E_g\propto \lambda^{m}$, where the non-universal exponent $m$ actually does depend on $D$.
This is analogous to the behavior of the corresponding 1D system~\cite{denNijs} where in the
absence of DM term one has $E_g \approx \lambda $. 
The BSRG for three-site problem in 1D with $D=0$ gives $E_g\propto\lambda^{0.63}$.  
In 2D square lattice Fig.~\ref{LnEgn1} suggests that this exponent for $D=0$
is given as $E_g \approx \lambda^{0.4869}$. 
Note that the value of the exponent ($0.4869$) has finite size errors. 

\begin{figure}[t]
 	\centering
 	\includegraphics[width=0.85 \linewidth]{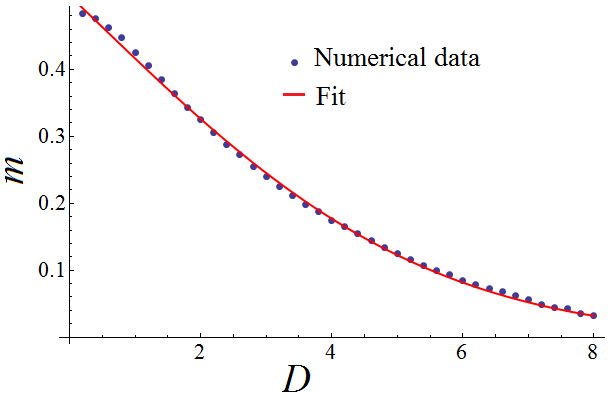}
 	\caption{(Color online) Exponent $m$ of $E_g\sim \lambda^m$ which determines
	how does the gap vanish as a function of $\lambda$. This exponent variers with
	$D$. These values are extracted from 8th level RG step which converges within 
	the precision of $10^{-5}$ }
 	\label{LnEgn}
\end{figure}

By turning on the DM interaction $D$ as can be seen in Fig.~\ref{gapXY} still the
gap vanishes as $\lambda$ approaches zero. To quantify this, we repeat the above 
log-log analysis for various values of $D$, and extracting the corresponding exponent $m$
as a function of $D$, we obtain the set of data points in Fig.~\ref{LnEgn}.
As can be seen from Fig.~\ref{LnEgn} for larger $D$ the exponent becomes smaller. 
Using the following ansatz for the fit,
\begin{eqnarray}
m=  \exp(\alpha D^2 + \beta D + \gamma),
\end{eqnarray}
gives, $\alpha=-0.02042 \pm 0.00106 $, $\beta=-0.1828 \pm 0.00523 $ and $\gamma=-0.6732 \pm 0.00499 $.

\subsection{Analysis of the concurrence}
So far we have established that for any $D$, the $\lambda=0$ repulsive line is a gapless line.
This is consistent with a picture of underlying phase coherent super-fluid, albeit not limited
to $D=0$, but also valid for nonzero values of $D$. 
The value of $D$ only affects the exponent $m$ that determines how fast the gap vanishes. 
Its repulsive nature indicates some form of instability towards a gapped state. Both 
positive and negative $\lambda$ sides are gapped states. Is the gap closing at $\lambda=0$ line
a topological phase transition?
In the $\lambda=0$ (isotropic) XY model, the non-analytic value of GMC is suggested as and 
indicator of the spin fluid phase in the 2D system \cite{Usman}.  
\begin{figure}[t]
	\centering
	\includegraphics[width=0.800 \linewidth]{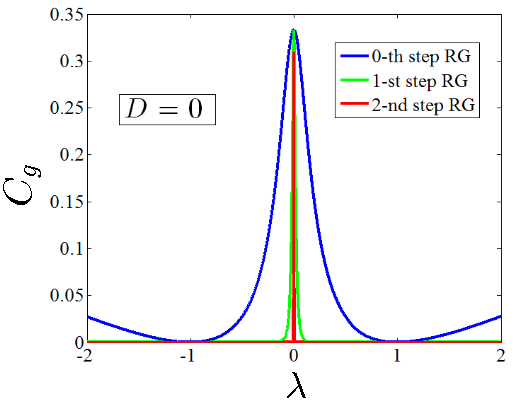}
	\caption{(Color online) GMC as a function of anisotropy parameter $\lambda$ in the absence 
	of DM interaction in different RG steps. The non-vanishing GMC indicates that the gap closing
	at $\lambda=0$ is a topological phase transition.}
	\label{GMC1}
\end{figure}
\begin{figure*}
	\centering
	\includegraphics[width=0.8500 \linewidth]{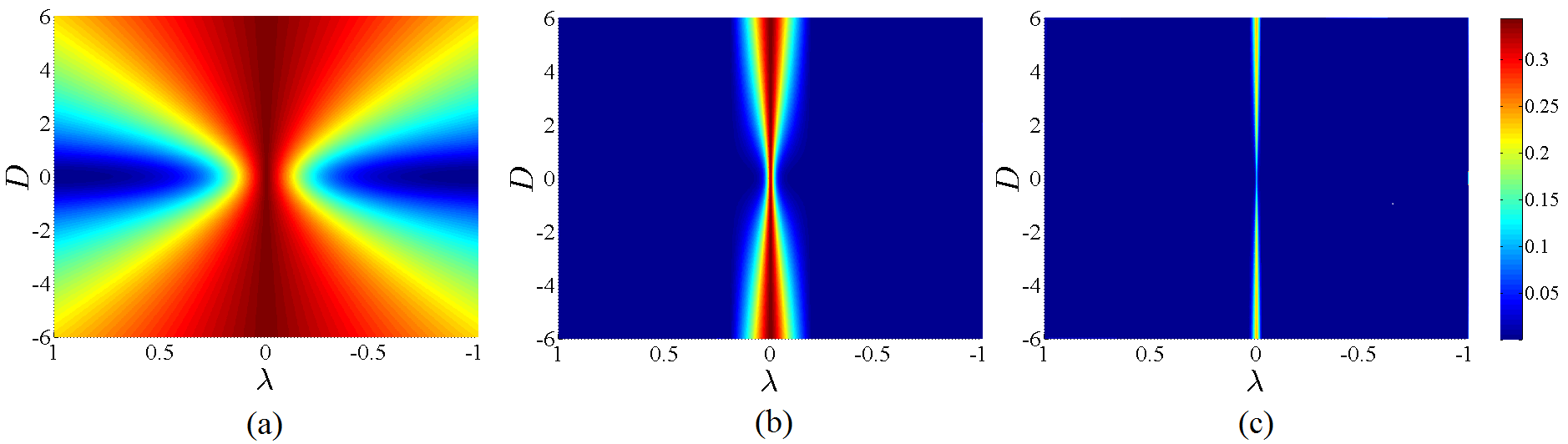}
	\caption{(Color online) Intensity map of GMC in the $(\lambda,D)$ plane for various RG steps:
		(a) 0-th RG step, (b) 1-st RG step and (c) 2-nd RG step. In the all values of DM interaction at the 
		nontrivial point of $\lambda =0$ the GMC shows non-analytic behavior.
	}
	\label{CgAll}
\end{figure*}
In Fig.~\ref{GMC1} we have plotted the GMC versus anisotropy parameter $\lambda$ for the $D=0$ case. 
As can be seen by repeating the RG steps, the convergence can be attained very quickly, and the GMC
at $\lambda=0$ becomes non-analytic. This suggest that the gap closing at $\lambda=0$ line
is a topological phase transition~\cite{SCI2017}. 
A nice feature of the above plot is the vanishing of GMS at $\lambda=\pm 1$ which 
corresponds to Ising-Kitaev limit polarized along $\hat x$ or $\hat y$ directions. 
For such a product state the entanglement must be zero. 

To put the above picture in a global perspective, in Fig.~\ref{CgAll} we plot
intensity profile of GMC at first two steps of the RG process. This figure suggests that
at the $\lambda=0$ line the gap-closing is accompanied by a change of topology~\cite{SCI2017}.

\section{Summary and discussion}\label{Discuss}
The phase portrait of anisotropy 2DQXY model with DM interaction in Fig.~\ref{platform} 
indicates that the DM interaction is irrelevant in the infrared limit. The $\lambda=0$ line
is a gapless line that separates two gapped states for positive and negative $\lambda$. 
The analysis of concurrence in Fig.~\ref{CgAll} suggest that the gap-closing transition at
$\lambda=0$ is a topological phase transition~\cite{SCI2017}.
In the bosonic language, the gapless state at $\lambda=0$ corresponds
to a super-fluid phase of underlying bosons~\cite{Rigol2012,SachdevBook}, and vanishing of the gap
can be attributed to the soft phase fluctuations of a super-fluid~\cite{Sandvik1999}.
There are two ways to destroy the long range order in the phase variable:
The well known way is by the BKT mechanism, i.e. the proliferation of vortices at elevated temperatures.

The second way to gap the super-fluid state 
is to stay at zero temperature but turn on the anisotropy $\lambda$. 
According to present study, 
as long as anisotropy $\lambda$ stays at zero, the DM interaction does not help
with gapping the state. Having established that $\lambda\ne0$ generates a gapped state
for any $D$, the question is, what kind of gapped state is it? Is it topologically
trivial or non-trivial? Fermionic representation of the problem in terms of Jordan-Wigner
fermions coupled with the Chern-Simons gauge fields~\cite{Fradkin1989,Fradkin1994,FradkinBook,Fradkin2014} 
suggests that the gapped state
is a topological superconductor~\cite{Kamenov2017}. The super-fluid picture at $\lambda=0$ (in the bosonic language)
corresponds to a liquid of JW fermions coupled with CS gauge fields in the fermionic picture. 
In the fermionic language, the anisotropy parameter $\lambda$ triggers a superconducting 
pairing instability in the Fermi sear of JW fermions
leading to a topologically non-trivial superconducting state of JW fermions~\cite{Kamenov2017}.

In our RG picture this can be understood as follows:
Deep in the gapped phase, at the Ising-Kitaev fixed points, $\lambda=+(-)1$ the long distance
behavior of the system is equivalent to a simple 2D Ising model polarized along $\hat x~(\hat y)$ direction. 
The ground state at these fixed points is factorizable and this explains why in Fig.~\ref{GMC1}
the entanglement indicator at all RG steps gives zero. This means that at the Ising-Kitaev fixed point
the Hamiltonian is given in terms of entirely commuting variables, and hence it has become
purely classical (hence zero entanglement). The fact that entanglement at every RG step (i.e. for
every system size) in Fig.~\ref{GMC1} is zero, already indicates that it has been protected 
by some sort of topology, and therefore the resulting Ising degeneracy can be interpreted 
in a dual picture as topological degeneracy~\cite{Fradkin2017}. 
At these fixed points the resulting classical 2D Ising model translates via celebrated Lieb-Schultz-Mattis mapping~\cite{mattisBook}
to a one-dimensional p-wave superconductor in modern terms. This is nothing but the well known Kitaev model of
a topological superconductor. Therefore the ground states at the fixed points $\lambda=\pm1$ is entitled to 
a winding number. 
Now moving slightly away from these fixed points and deforming the Hamiltonian in such a way
that it ultimately returns to the fixed points upon enlarging the length scale, the topological number
does not change, as there is no gap-closing as long as one does not hit the $\lambda=0$ repeller line. 
Therefore our real space RG is consistent with a non-trivial topological charge
for the gapped states at $\lambda\ne 0$. 

To summarize, we have considered the quantum XY model in 2D square lattice in the 
presence of DM interaction. The symmetry of problem allows us to obtain 
analytical expressions for the ground state doublet of this system which then
enables us to set up a real space block spin RG. The DM interaction turns out
to be irrelevant at long wave-lengths. The RG flow consists in a gapless repulsive
$\lambda=0$ line, and two attractive $(\lambda=\pm1,D=0)$ points corresponding to 
Ising-Kitaev limit. Non-analyticity of concurrence shows that the phase transition at
$\lambda=0$ is of topological nature~\cite{SCI2017}.  
The Ising Kitaev-limit enables us to assign a topological charge
to the gapped phases at $\lambda\ne 0$. These features are very similar
to corresponding 1D system~\cite{SAJ2017} and in agreement with results of studies
based on JW fermions coupled with CS gauge fields~\cite{Kamenov2017}. 

\section*{ACKNOWLEDGMENT}
SAJ appreciates financial supports by Alexander von
Humboldt fellowship for experienced researchers
\appendix
\section{Details of exact diagonalization of selected cluster in square lattice}
In this appendix, details of the exact diagonalization for 
selected cluster in square lattice are presented. To reduce 
the dimension of ensuing matrix we employ group theory method. 
To obtain the eigenvalues (Eq.~\ref{Eig}) and eigen-states 
(Eq.~\ref{Eigv1} and \ref{eigv2}) first we consider the possible states 
of spin-1/2 system $2^5$ in cluster $P$. Each state of the cluster is in 
the following form,
\begin{eqnarray}
| \alpha_i \rangle = | \sigma_4, \sigma_3 ,\sigma_2 ,\sigma_1 ,\sigma_0 \rangle 
\end{eqnarray}
where $i=1~...~32$ and $\sigma$ present the two possible values $\uparrow \downarrow$ 
in Fig.~\ref{Cluster} . The basis in this 32 dimensional Hilbert space are as (for brevity in representation 
of basis states we drop $|\rangle$),
\begin{eqnarray}
&|\alpha_1 \rangle = \uparrow \uparrow \uparrow \uparrow \uparrow,
|\alpha_2 \rangle = \uparrow \uparrow \uparrow \uparrow \downarrow,
|\alpha_3\rangle = \uparrow \uparrow \uparrow \downarrow \uparrow,
|\alpha_4 \rangle = \uparrow \uparrow \downarrow \uparrow \uparrow, \nonumber\\
&|\alpha_5 \rangle = \uparrow \downarrow\uparrow \uparrow \uparrow,
|\alpha_6 \rangle = \downarrow \uparrow \uparrow \uparrow \uparrow,
|\alpha_7 \rangle = \downarrow \uparrow \uparrow \uparrow \downarrow,
|\alpha_8 \rangle = \downarrow \uparrow \uparrow \downarrow \downarrow, \nonumber\\
&|\alpha_9 \rangle =\downarrow \uparrow \downarrow \uparrow \downarrow,
|\alpha_{10} \rangle  =\downarrow \downarrow \uparrow \uparrow \downarrow,
|\alpha_{11} \rangle  =\uparrow \uparrow \uparrow \downarrow \downarrow,
|\alpha_{12} \rangle  =\uparrow \uparrow \downarrow \downarrow \uparrow, \nonumber\\
&|\alpha_{13} \rangle  =\uparrow \downarrow \downarrow \uparrow \uparrow,
|\alpha_{14} \rangle  =\uparrow \downarrow \uparrow \uparrow \downarrow,
|\alpha_{15} \rangle  =\uparrow \downarrow \uparrow \downarrow \uparrow,
|\alpha_{16} \rangle   =\uparrow \uparrow \downarrow \uparrow \downarrow, \nonumber\\
&|\alpha_{17} \rangle = \downarrow \downarrow \downarrow \downarrow \uparrow,
|\alpha_{18} \rangle =\downarrow \downarrow \downarrow \uparrow \downarrow,
|\alpha_{19} \rangle =\downarrow \downarrow \uparrow \downarrow \downarrow,
|\alpha_{20} \rangle =\downarrow \uparrow \downarrow \downarrow \downarrow,\nonumber\\
&|\alpha_{21} \rangle =\uparrow \downarrow \downarrow \downarrow \downarrow,
|\alpha_{22} \rangle =\uparrow \downarrow \downarrow \downarrow \uparrow,
|\alpha_{23} \rangle =\uparrow\downarrow \downarrow \uparrow \downarrow,
|\alpha_{24} \rangle =\uparrow \downarrow \uparrow \downarrow \downarrow,\nonumber\\
&| \alpha_{25} \rangle = \uparrow \uparrow \downarrow \downarrow \downarrow,
| \alpha_{26} \rangle = \downarrow \downarrow \downarrow \uparrow \uparrow,
| \alpha_{27} \rangle = \downarrow \downarrow \uparrow \uparrow \downarrow,
| \alpha_{28} \rangle =\downarrow \uparrow \uparrow \downarrow \downarrow, \nonumber\\
&| \alpha_{29} \rangle = \downarrow \uparrow \downarrow \downarrow \uparrow,
| \alpha_{30} \rangle = \downarrow \uparrow \downarrow \uparrow \downarrow,
| \alpha_{31} \rangle = \downarrow \downarrow \uparrow \downarrow \uparrow,
| \alpha_{32} \rangle =  \downarrow \downarrow \downarrow \downarrow \downarrow,\nonumber\\
\end{eqnarray}
Now we proceed calculations by employing symmetry consideration to reduce 
32 dimensional Hilbert space to smaller blocks in matrix representation. The 
+ shape of cluster in Fig.~\ref{Cluster}. is invariant under rotations by $\frac{\pi}{2}$ 
which is denoted by $\mathcal{C}$ and then the rotation group is given by 
$\{\mathcal{C}^0 , \mathcal{C}^1,\mathcal{C}^2,\mathcal{C}^3 \}$. 
The $\mathcal{C}$ operates on the site labels as,
\begin{eqnarray}
\mathcal{C} = \left\{ \begin{matrix}
1\to 2  \\
2\to 3  \\
3\to 4  \\
4\to 1  \\
\end{matrix} \right.\
\end{eqnarray}
By successive operation of $\mathcal{C}$ on a one state for e.i. $|\alpha_3 \rangle $, 
the following pattern is obtained,
\begin{eqnarray}
|3\rangle\xrightarrow{\mathcal{C}}|4\rangle\xrightarrow{\mathcal{C}}|5\rangle\xrightarrow{\mathcal{C}}|6\rangle\xrightarrow{\mathcal{C}}|3\rangle
\end{eqnarray}
which is the concise representation of 
\begin{eqnarray}
\mathcal{C}^0 |3 \rangle =  |3 \rangle, \nonumber\\
\mathcal{C}^1 |3 \rangle =  |4 \rangle, \nonumber\\
\mathcal{C}^2 |3 \rangle =  |5 \rangle, \nonumber\\
\mathcal{C}^3 |3 \rangle =  |6 \rangle, \nonumber\\
\label{Roo}
\end{eqnarray}
According to projection theorem in group theory we 
construct the symmetry adopted state in representation 
which is labeled by $n$ from an arbitrary state $| \phi \rangle$ 
\begin{eqnarray}
|\psi^{(n)}\rangle \sim \left(\sum_{g} g \Gamma_n[g]\right)|\phi\rangle
\end{eqnarray}
where $g$ interprets the member of group and $\Gamma_n[g]$ denotes 
the $n$-th irreducible representation for element $g$ in the group. 
Our case is a rotation group and the irreducible representations 
of the cyclic group are tagged by means of three (angular momentum) 
$n=0,~\pm1$. These are presented by $\{\omega^0,~\omega^n,~\omega^{2n},~\omega^{3n} \}$ 
where $\omega = \exp(i\pi/2)$. The $\Gamma_n(\mathcal{C}^p) = \omega^{pn}$ is the 
well-set representation of above cyclic group. A symmetry adopted 
state build from e.i. $|3 \rangle$ is,
\begin{eqnarray}
\left(\mathcal{C}^0\omega^0 + \mathcal{C}^1\omega^n + \mathcal{C}^2\omega^{2n} +\mathcal{C}^3\omega^{3n} \right)|3\rangle
\end{eqnarray}
where by applying Eq.~\ref{Roo}, the obtained state is as,
\begin{eqnarray}
|3\rangle + \omega^n |4\rangle + \omega^{2n} |5\rangle + \omega^{3n} |6\rangle
\end{eqnarray}
 with $n$ is the angular momentum.\
By applying the same symmetry to every other states we 
obtain,
\begin{eqnarray}
&|7\rangle\xrightarrow{\mathcal{C}}|11\rangle\xrightarrow{\mathcal{C}}|14\rangle\xrightarrow{\mathcal{C}}|16\rangle\xrightarrow{\mathcal{C}}|7\rangle \nonumber\\
&|8\rangle\xrightarrow{\mathcal{C}}|10\rangle\xrightarrow{\mathcal{C}}|12\rangle\xrightarrow{\mathcal{C}}|13\rangle\xrightarrow{\mathcal{C}}|8\rangle \nonumber\\
&|18\rangle\xrightarrow{\mathcal{C}}|19\rangle\xrightarrow{\mathcal{C}}|20\rangle\xrightarrow{\mathcal{C}}|21\rangle\xrightarrow{\mathcal{C}}|18\rangle \nonumber\\
&|22\rangle\xrightarrow{\mathcal{C}}|26\rangle\xrightarrow{\mathcal{C}}|29\rangle\xrightarrow{\mathcal{C}}|31\rangle\xrightarrow{\mathcal{C}}|22\rangle \nonumber\\
&|23\rangle\xrightarrow{\mathcal{C}}|25\rangle\xrightarrow{\mathcal{C}}|27\rangle\xrightarrow{\mathcal{C}}|28\rangle\xrightarrow{\mathcal{C}}|23\rangle \nonumber\\
&|9\rangle\xrightarrow{C}|15\rangle\xrightarrow{C}|9\rangle\nonumber\\
&|24\rangle\xrightarrow{C}|30\rangle\xrightarrow{C}|24\rangle\nonumber\\
&|1\rangle\xrightarrow{C}|1\rangle,~~~~~|2\rangle\xrightarrow{C}|2\rangle\nonumber\\
&|17\rangle\xrightarrow{C}|17\rangle,~~~~~|32\rangle\xrightarrow{C}|32\rangle\nonumber\\
\end{eqnarray}
The normalized states are as,
\begin{eqnarray}
&|\phi_1\rangle = |\alpha_1 \rangle, \nonumber\\
&|\phi_2\rangle = |\alpha_2 \rangle, \nonumber\\
&|\phi_3\rangle = \frac{1}{2}\left(|\alpha_3 \rangle + |\alpha_4 \rangle + |\alpha_5 \rangle +|\alpha_6 \rangle\right) , \nonumber\\
&|\phi_4\rangle = \frac{1}{2}\left(|\alpha_7 \rangle + |\alpha_{11} \rangle + |\alpha_{14} \rangle +|\alpha_{16} \rangle\right) , \nonumber\\
&|\phi_5\rangle = \frac{1}{2}\left(|\alpha_8 \rangle + |\alpha_{10} \rangle + |\alpha_{12} \rangle +|\alpha_{13} \rangle\right) , \nonumber\\
&|\phi_6\rangle = \frac{1}{\sqrt{2}}\left(|\alpha_9 \rangle + |\alpha_{15} \rangle \right) , \nonumber\\
&|\phi_7\rangle = \frac{1}{2}\left(|\alpha_{18} \rangle + |\alpha_{19} \rangle + |\alpha_{20} \rangle +|\alpha_{21} \rangle\right) , \nonumber\\
&|\phi_8\rangle = \frac{1}{2}\left(|\alpha_{22} \rangle + |\alpha_{26} \rangle + |\alpha_{29} \rangle +|\alpha_{31} \rangle\right) , \nonumber\\
&|\phi_9\rangle = \frac{1}{2}\left(|\alpha_{23} \rangle + |\alpha_{25} \rangle + |\alpha_{27} \rangle +|\alpha_{28} \rangle\right) , \nonumber\\
&|\phi_{10}\rangle = \frac{1}{\sqrt{2}}\left(|\alpha_{24} \rangle + |\alpha_{30} \rangle \right) , \nonumber\\
&|\phi_{11}\rangle = |\alpha_{17} \rangle, \nonumber\\
&|\phi_{12}\rangle = |\alpha_{32} \rangle. \nonumber\\
\end{eqnarray}
The same approach will lead to normalized state at $n=+1$ sector. Due 
to the time reversal symmetry the $n=-1$ sector has identical spectrum. 
The $n=+1$ sector normalized states are,
\begin{eqnarray}
&|\chi_1\rangle = \frac{1}{2}\left(|\alpha_3 \rangle + \omega |\alpha_4 \rangle + \omega^2|\alpha_5 \rangle +\omega^3|\alpha_6 \rangle\right) , \nonumber\\
&|\chi_2\rangle = \frac{1}{2}\left(|\alpha_7 \rangle + \omega|\alpha_{11} \rangle + \omega^2|\alpha_{14} \rangle +\omega^3|\alpha_{16} \rangle\right) , \nonumber\\
&|\chi_3\rangle = \frac{1}{2}\left(|\alpha_8 \rangle + \omega|\alpha_{10} \rangle +\omega^2 |\alpha_{12} \rangle +\omega^3|\alpha_{13} \rangle\right) , \nonumber\\
&|\chi_4\rangle = \frac{1}{2}\left(|\alpha_{18} \rangle + \omega|\alpha_{19} \rangle + \omega^2|\alpha_{20} \rangle +\omega^3|\alpha_{21} \rangle\right) , \nonumber\\
&|\chi_5\rangle = \frac{1}{2}\left(|\alpha_{22} \rangle + \omega|\alpha_{26} \rangle + \omega^2|\alpha_{29} \rangle +\omega^3|\alpha_{31} \rangle\right) , \nonumber\\
&|\chi_6\rangle = \frac{1}{2}\left(|\alpha_{23} \rangle + \omega|\alpha_{25} \rangle +\omega^2 |\alpha_{27} \rangle +\omega^3|\alpha_{28} \rangle\right) , \nonumber\\
\end{eqnarray}
It should be noted that the other symmetry such as parity symmetry 
in the selected cluster is in the heart of the rotation symmetry. 
The other operator that we introduced is 
\begin{eqnarray}
\zeta = \prod_{i} \sigma^z_i
\end{eqnarray}
which operates as a constant of motion. Consider one arbitrary 
state with arrangements of spins of up and down. The operation 
of XY Hamiltonian on a selected arrangements does not change the 
value of $q$. The reason is that in the presence of the two 
consecutive $\sigma^x$ or $\sigma^y$ operator the total number 
of spin flip is even. This operator acts on the 32 basis of cluster 
and breaks it in two family with $\zeta=+1$ which consist of
\begin{eqnarray}
|\alpha_1 \rangle , |\alpha_7 \rangle , |\alpha_8 \rangle, |\alpha_9 \rangle \nonumber\\
|\alpha_{10} \rangle, |\alpha_{11} \rangle , |\alpha_{12} \rangle,|\alpha_{13} \rangle \nonumber\\
|\alpha_{14} \rangle , |\alpha_{15} \rangle, |\alpha_{16} \rangle, |\alpha_{17} \rangle \nonumber\\
|\alpha_{18} \rangle,|\alpha_{19} \rangle,|\alpha_{20} \rangle,|\alpha_{21} \rangle
\end{eqnarray}
and $\zeta=-1$ 
\begin{eqnarray}
|\alpha_2 \rangle , |\alpha_3 \rangle , |\alpha_4 \rangle, |\alpha_5 \rangle \nonumber\\
|\alpha_{6} \rangle, |\alpha_{22} \rangle , |\alpha_{23} \rangle,|\alpha_{24} \rangle \nonumber\\
|\alpha_{25} \rangle , |\alpha_{26} \rangle, |\alpha_{27} \rangle, |\alpha_{28} \rangle \nonumber\\
|\alpha_{29} \rangle,|\alpha_{30} \rangle,|\alpha_{31} \rangle,|\alpha_{32} \rangle.
\end{eqnarray}
By considering all the symmetries and constant of motion, it 
is possible to diagonalize Hamiltonian analytically for obtain the 
ground state and energy bands. For e.i. in $n=0$ sector the Hamiltonian 
of the system by considering above symmetries in $\zeta=1$ reduced to
\begin{widetext}
\begin{eqnarray*}
 H =
\left[
\begin{matrix}
0 & 0 & 0 & 4J\lambda & 0 & 0 \\
0 & 0 & 4J(1+iD) & 0 & 0 & 0 \\
0 & 4J(1 - iD) & 0& 0 & 4J\lambda & 2\sqrt{2}J\lambda \\
 4J\lambda 0 & 0 & 0 & 0& 4J(1+iD) & 2\sqrt{2}J(1 + iD) \\
0 & 0 & 4J\lambda & 4J(1-iD) & 0 & 0\\
0 & 0 & 2\sqrt{2}J\lambda& 2\sqrt{2}J(1 - iD) & 0 & 0\\
\end{matrix}
\right]
\end{eqnarray*}
where the eigenvalues of above matrix of Hamiltonian are
\begin{eqnarray}
& e_0 = -2J \sqrt{5 (1 +D^2)+ 5 \lambda^2 + \eta}, \nonumber\\
& e_1 = -2J \sqrt{5 (1 +D^2)+ 5 \lambda^2 - \eta}, \nonumber\\
& e_2 = e_3 = 0, \nonumber\\
&e_4 = 2J \sqrt{5 (1 +D^2)+ 5 \lambda^2 - \eta}, \nonumber\\ 
&e_5 = 2J \sqrt{5 (1 +D^2)+ 5 \lambda^2 + \eta}, \nonumber\\ 
\end{eqnarray}
in which 
\begin{eqnarray}
\eta = \sqrt{\lambda^4 + 34 \lambda^2 (1 + D^2) + (1 + D^2)^2}
\end{eqnarray}
and the $e_0$ is the ground state.
\end{widetext}

\bibliographystyle{apsrev4-1}
\bibliography{Refs}
\end{document}